# Extraction of hidden information by efficient community detection in networks

Juyong Lee[1], Steven P Gross[2,1] and Jooyoung Lee[1*]

[1]*School of Computational Sciences, Korea Institute for Advanced Study, Seoul, Korea*

[2] *Department of Developmental and Cell biology, University of California, Irvine, USA,*

*E-mail: juyong@kias.re.kr, Telephone: 82-2-958-3817, Fax: 82-2-958-3890*

*E-mail:sgross@uci.edu, Telephone: 1-949-824-3159, Fax: 1-949-824-4709*

*E-mail: jlee@kias.re.kr, Telephone: 82-2-958-3731, Fax: 82-2-958-3890*

***Summary***

Currently, we are overwhelmed by a deluge of experimental data, and network physics has the potential to become an invaluable method to increase our understanding of large interacting datasets. However, this potential is often unrealized for two reasons: uncovering the hidden community structure of a network—known as community detection—is difficult, and further, even if one has an idea of this community structure, it is not *a priori* obvious how to efficiently use this information. Here, to address both of these issues, we, first, identify optimal community structure of given networks in terms of modularity by utilizing a recently introduced community detection method. Second, we develop an approach to use this community information to extract hidden information from a network. When applied to a protein-protein interaction network, the proposed method outperforms current state-of-the-art methods that use only the local information of a network. The method is generally applicable to networks from many areas.

## *Introduction*

Recently, network physics has been used to provide a framework for investigating the structure and dynamics of complex systems in various fields[1-3]. Typical real-world networks adopt hierarchical structures, including communities composed of more densely inter-connected subgraphs. Uncovering the hidden community structure of a network, known as community detection or module detection, has been a subject of active research in mathematical, social and physical sciences[4,5]. Nodes in a community likely share common characteristics, so proper community detection in principle allows extraction of hidden information from the network, without additional *a priori* knowledge. Biologically, a group of proteins or genes in the same community often coincides with known functional modules and/or protein complexes[6,7]. Similar functional correspondence is also observed in gene co-expression networks[8]. While such communities are thus in principle invaluable, to date the utility of this information has been limited both by the difficulty in accurately detecting the communities, and also, by lack of formal optimized ways to incorporate this community information into a generalized approach to extract hidden information. For this reason, module-assisted methods have so far been inferior to simple neighbor-assisted approaches for function prediction of proteins from a protein-protein interaction (PPI) network, one of the ultimate testing grounds of community detection[5, 9-11].

Among various definitions of a community, modularity maximization is most widely used[5,12]. The modularity, Q, measures the relative density of intra-community connectivity, compared to a randomly re-wired counterpart with the same degree of nodes. Maximization of Q recasts the community detection problem into a global optimization problem. As the network size increases, the complexity of Q-maximization increases more rapidly than exponential growth, so simple enumeration is impractical. Until now, rather than attacking the difficult problem of developing methods to find the optimal solution, many fast heuristic methods have been employed. Therefore, despite its popularity, little is known about the utility of a community detected via Q-maximization. A fundamental assumption is that the quality of community structure correlates with its Q, and that from higher quality communities one can ultimately derive more accurate insights, but the actual validity of this hypothesis in the context of real-world applications has rarely been shown.

Recently, we proposed a new community detection method based on Q-maximization, using the global optimization technique called conformational space annealing (CSA)[13,14]. CSA has been quite efficient in solving difficult combinatorial optimization problems including recent success in protein structure prediction[15-18]. Compared to simulated annealing (SA)— considered to be the most accurate method so far[19-21]—applying CSA to community detection provides higher Q partitioning, is much faster, and solutions converge, displaying far less variability. In this letter, we show that achieving higher Q partitioning does indeed unveil additional hidden information. The method is generally applicable to any networks, and tested on a PPI network where community-based methods have not been much successful[11].

## *Method*

Once the community structure of a network is identified, the significance of the generated sub-networks can be tested by what is called functional enrichment, which measures the degree of flocking together among birds of a feather. The efficiency of CSA-based Q-maximization has been already demonstrated on popular real world networks[22]. Here, three biological networks are tested: two metabolic networks of *T. pallidum* and *E. coli*, and the PPI network of *Saccharomyces cerevisiae*. For the *T. pallidum* study, we generated the network according to the previous study[23]. The other two networks and related meta data were kindly provided by Ahn, Bagrow and Lehmann[23]. The results from these three networks are summarized in Fig. 1.

We observe that CSA-generated community structures are of higher modularity than SA-generated ones, as judged by Q values. How useful are the detected communities? In previous studies[2, 20, 24], it has not yet been clearly shown that whether a higher Q partition is more meaningful than low Q partitions. To determine the effectiveness of the partitioning, we used two measures. First, we looked at the number of identified enriched functional clusters, i.e. those that belong to a particular community in a significantly non-random fashion. In all cases, relative to SA, the CSA-generated community structure has a larger number of enriched functional clusters (Fig. 1, left), suggesting that the sub-groups generated by CSA were more meaningful than those by SA. We emphasize that the result is insensitive to the variation the P value threshold (not shown).

We also examined the extent of common features shared between nodes in the same community defined as $E = \frac{\sum_c n_c \mu_c}{\sum_c n_c}$ where $\mu_c$ is the average feature similarity between pairs in community $c$ containing $n_c$ nodes. The feature similarity between two nodes is defined as unity if they share a feature. In general we found positive correlation between increased modularity Q and increased quality $E$ (Fig. 1, right). Based on these results, the communities detected by CSA, with increased modularity, are likely to capture useful 'hidden' information.

One important feature of CSA-mediated community detection is that it is much more reproducible. This can be seen in Fig. 1; for the smaller *T. pallidum* and *E. coli* networks, CSA repeatedly found a single solution (red circle), better than any of the many solutions (blue crosses) found by SA. Similarly, for the more challenging *S. cerevisiae* network, the multiple solutions found by CSA (red circles) are all better than the multiple solutions found by SA (blue crosses).

### *Using community information to improve protein function prediction*

Protein function prediction is one of the most important issues in the current post-genomic era. Biological interaction networks are modular, so finding the community structure of a PPI network has been regarded as a promising approach to improve protein function prediction over existing methods. Surprisingly, however, attempts to incorporate community structure for function prediction have been somewhat limited and not very successful, especially compared to simple neighbor-assisted methods[9-11, 25, 26]. We suggest that this failure of community-based methods reflects both a failure in correct detection of the underlying communities (see above), combined with an over-simplified way of utilizing community information (see next).

In most community-based approaches to date, if a function from a community is signified by a low P value, the function is assigned to all nodes in the community. This approach is likely to induce large false-positives, leading to low prediction accuracy, and makes such approaches of limited utility[10, 11].

As an alternative to such a low-resolution assignment of function, we employed a random forest (RF)[27] machine learning technique. We applied it on the yeast PPI network containing 2729 proteins by carrying out leave-one-out cross-validation with GO annotations[28]. That is, for each protein, we assumed that its function(s) were unknown. We then made predictions of its function(s) based on only a set of input features generated from the network (including e.g. which communities its neighbors were in, their functions, etc.). By completely deleting a node/protein and considering the network and GO annotations from the other 2728 proteins, an RF consisting of 500 classification trees was trained to maximize the function prediction for 2728 proteins. The trained machine was then used to predict the function of the deleted protein. This procedure was repeated separately for all 2729 proteins. Community structure was used as one of the inputs into the prediction by including it in the feature vector for RF. To determine the importance of this information, this feature vector was constructed with and without community information. For each node $i$, we consider each function $f$ from its neighbors as a candidate function for $i$, and we want to design a machine to decide the adequacy of $f$. For each $i$, we calculate 11 features, 5 using only neighbor information, 5 by combining community information with the neighbor information and one background information, the fraction of proteins with function $f$.

The first 5 features are 1) number of neighbors, 2) number of neighbors with $f$, 3) fraction of neighbors with $f$, 4) rank of $f$ frequency among all functions from neighbors and 5) P-value of $f$ calculated from the neighbors against all nodes. The next 5 features are calculated from neighbors and the community where $i$ belongs; 6) number of neighbors in the same community of $i$, 7) ratio of 6) to all neighbors, 8) fraction of 6) with $f$, 9) P-value of $f$ in the community of $i$ against all nodes, 10) P-value of $f$ based on 6).

While applying this prediction scheme, we are confronted with two technical issues, long computational time due to the large number of training examples and the unbalanced ratio between small positive and large negative examples. To alleviate these problems, we took all positive examples and only the same number of randomly selected negative examples for training.

For comparison, we also carried out all currently available outstanding methods including majority voting[25], neighborhood enrichment[10, 11] and two Markovian random field (MRF) methods[29, 30] which all utilize only local information such as the ranking of functional frequency present in the neighboring proteins. These methods are currently the state of the art in the field of protein function prediction[31].

When making predictions, two aspects are important: the fraction of correct prediction (precision), and the fraction of total retrieval (recall). Obviously, there is a trade-off: one can make only those few predictions about which one has high certainty, or one can strive to make more predictions, at the cost of increased error. In practice, this can be summarized by a Precision-Recall curve, and the integrated area under this curve (AUC)[32] is a quantitative measure of a particular method.

In Table 2, we use this AUC metric to compare the efficiency of various methods. MRF methods are currently considered to be the most efficient, and indeed MRF by Karaoz et al.[30] performs better than the other existing methods in our hands. However, for all three GO domains (biological process (BP), cellular component (CC) and molecular function (MF)), our new implementation of the RF method (RF-comm-CSA), utilizing the best CSA solution performed the best overall. While not as good as RF-comm-CSA, the AUC values of RF-local without community information were still better than MRF on average. Importantly, when the community information was properly used, the improvement of RF-comm-CSA over MRF by Karaoz et al. was 15.9%/4.4%/8.7% for BP/CC/MF.

Armed with RF-local and RF-comm, as well as better and worse partitioning (from application of CSA vs. SA), we were now able for the first time to address the question: 'when is community information useful in protein function prediction, and what aspects of the community structure are important?' RF-local performed slightly better than MRF by Karaoz et al.[30], by 2.8% on average for three GO domains. Thus, using RF with only local information is already quite good, better than the current gold standard. When we use RF-comm, as input we give it either the SA- or CSA-determined community structure. There is indeed an improvement of RF-comm over RF-local for both SA- or CSA-determined solutions (Table 2), answering part of the question: properly used, community information is valuable.

Intriguingly, the improvement of RF-comm-CSA and that of RF-comm-SA was roughly similar to each other for prediction of cellular component (CC) information (7.8% improvement in AUC using CSA, and 6.9% improvement using SA), but the improvement of RF-comm-CSA over SA was more pronounced for both biological process (BP) prediction (3.9% for CSA *vs.* 0.9% for SA) and for Molecular Function (MF) prediction (8.6% for CSA *vs.* 2.2% for SA). Thus, the additional information in the CSA-determined community was useful for both BP and MF prediction, but particularly important for MF prediction. Why is this so? We reasoned that there are likely some 'easy' aspects of community detection that both CSA and SA do reasonably well, that are useful for CC, and some harder aspects of community detection that CSA does better at than SA, and that matter for BP and especially MF.

In general, if nodes are close by, they are easy/likely to be together in a community, but the further apart two nodes are in a network, the harder it is to accurately place them in a community. Thus, we hypothesized that perhaps proximity is a relevant factor to consider, and the improvement in MF and to a lesser extent BP came from the fact that non-local information—as summarized by detected community structures— was useful in those cases. To test this idea, we calculated $p(n)$, the probability of sharing a common feature between two nodes separated by $n$ edges (see Fig. 2). We consider all possible pairs to obtain $p(n)$ for CC, BP, and MF, and observe that $p(n)$ for CC decays monotonically as a function $n$ while $p(n)$ for MF increases for $n>5$. $p(n)$ for BP is about constant for $n>3$. Thus, for CC, most common features are shared between small $n$ pairs and one might expect that the differences between two community structures are not crucial. However, for MF, large $n$ pairs contribute relatively more, and therefore details of community structures are important. This matches the improvement we see from incorporating community structure—when long-range relationships are not negligible, correct community partitioning capturing these relationships correctly can be especially useful. Thus, our data suggest that although community information is in general useful, accurate community structure is particularly important for prediction of molecular function and to a lesser extent biological processes, where non-local community information is relevant.

The above data suggests that long-range non-local data is particularly present for MF, and therefore may under specific conditions be a significant contributor to the accuracy of prediction. Intuitively, its utility will likely depend not only on how well such long-range information is captured (by appropriate community structure) but also how much local information there is to use in its place—if local information is lacking, non-local information should be particularly important. To test this idea, we calculated the improvement of AUC values separately, considering nodes within a range of a number of neighbors. That is, we considered separately nodes that had only a few local inputs (k<=3) to nodes with many local inputs (k >20). The result is shown in Table 2.

Community information was most useful in predicting functions of sparsely connected proteins with less than ten interacting proteins for MF. Not only does this conceptually make sense, it is encouraging from a practical point of view, since newly investigated unannotated proteins are likely to have a small number of edges/connections to other proteins.

## *Conclusion*

To the best of our knowledge, the approach used in this work is the first network-based method for successful protein function prediction utilizing community information which clearly outperforms existing best local-information based methods. Our results indicate that the community structure itself contains useful information, and by using a sophisticated data-mining technique we can extract this information. It is promising that with only computational efforts, without additional experimental information, the quality of prediction can be significantly enhanced.

Our approach, data-mining from global and local topological features of a network, can be a general framework for predicting hidden properties from social as well as biological networks. In particular, we believe that the conceptual advance clarifying when local vs. longer range community information is important—and how to approach such a question— will have ramifications for many disciplines where network physics is used.

**Acknowledgements**

The authors thank Y.-Y. Ahn, J.P. Bagrow and S. Lehmann for providing their networks and corresponding meta data for *E.coli* metabolic network and *S. Cerevisiae* PPI network. This work was supported by the National Research Foundation of Korea(NRF) grant funded by the Korea government(MEST) (No. 20120001222). We thank Korea Institute for Advanced Study for providing computing resources (KIAS Center for Advanced Computation Linux Cluster) for this work. We also like to acknowledge the support from the KISTI Supercomputing Center (KSC-2012-C3-01).

Correspondence and Requests for materials should be addressed to:  85 Hoegiro, Dongdaemun-gu, Seoul, 130-722, Republic of Korea. E-mail: jlee@kias.re.kr, Telephone: 82-2-958-3731, Fax: 82-2-958-3786.

## Tables

**Table 1. The efficiency of protein function prediction methods are summarized. RF-comm/neigh refers to the random forest method with/without community information. MRF refers to Markovian random field.**

| Methods | AUC | | |
|---|---|---|---|
| | **BP** | **CC** | **MF** |
| RF-comm-CSA (Q=0.7737) | 0.343 | 0.528 | 0.201 |
| RF-comm-SA (Q=0.7684) | 0.333 | 0.524 | 0.189 |
| RF-local | 0.330 | 0.490 | 0.185 |
| MRF by Karaoz et al. | 0.296 | 0.506 | 0.185 |
| MRF by Deng et al. | 0.266 | 0.436 | 0.165 |
| Neighborhood enrichment | 0.273 | 0.379 | 0.146 |
| Majority voting | 0.159 | 0.389 | 0.131 |

**Table 2 AUC values and relative improvements of prediction of molecular function by using community information are displayed. AUC values are calculated considering nodes with k in the range shown.**

| Number of neighbors (k) | Number of proteins | RF-comm-CSA | RF-local | Improvement (%) |
|---|---|---|---|---|
| k≤3 | 1358 | 0.114 | 0.103 | 10.7 |
| 3<k≤10 | 689 | 0.229 | 0.192 | 19.1 |
| 10<k≤20 | 343 | 0.340 | 0.334 | 1.8 |
| k>20 | 340 | 0.360 | 0.368 | -2.2 |

Improved protein function prediction achieved by community information

**Figures**

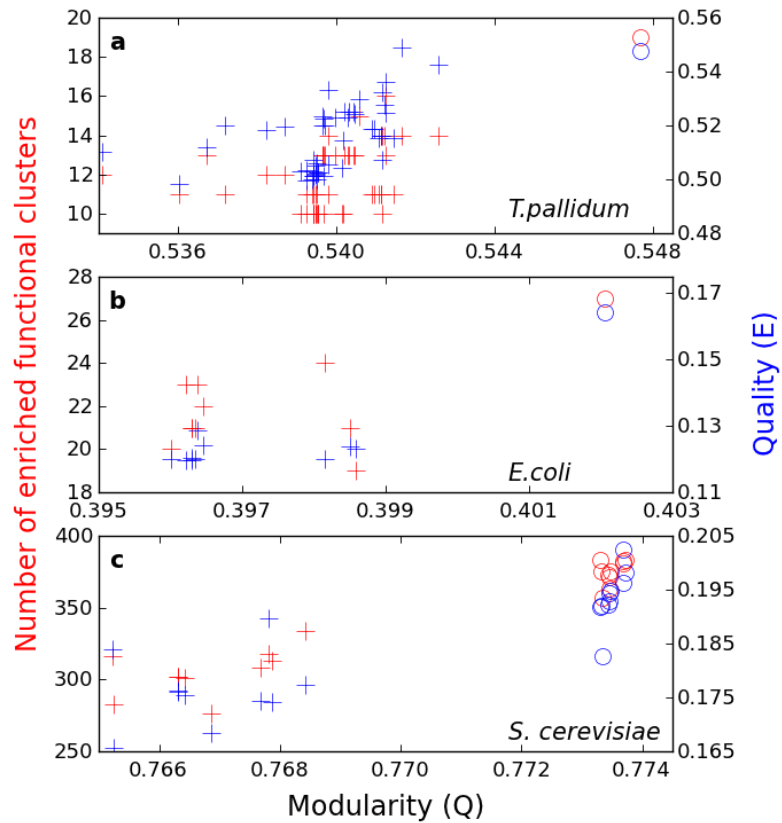

Figure 1 For three biological networks we display the relationship between modularity Q and two quality measures, the number of enriched functional clusters (left, red) and quality $E$ (right, blue). We used P value thresholds of $10^{-2}$, $10^{-4}$ and $10^{-7}$ for *T. pallidum, E. coli* and *S. cerevisiae*, respectively. For all three networks, CSA results (o) are showing higher Q values and better qualities in both measures than SA ones (+). Note that CSA runs all converged into identical solutions for metabolic networks of *T.pallidum* and *E.coli.*

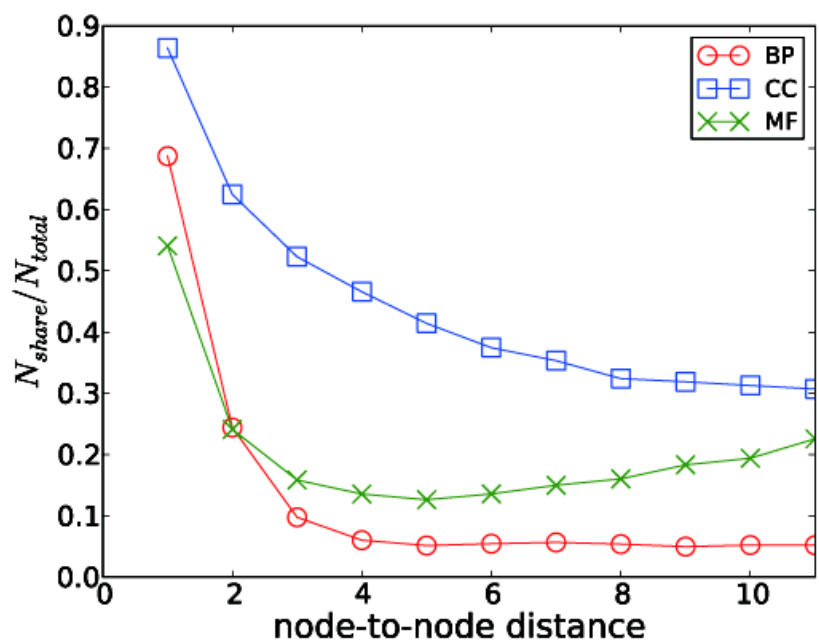

Figure 2 Fraction of protein pairs sharing a common GO term of biological process (red), cellular component (blue) and molecular function (green) domains is shown as a function of node-to-node distance in the PPI network.